\documentclass[twocolumn]{aastex61}
\usepackage{graphicx}
\usepackage{amsmath, amssymb, amsthm}
\usepackage{multirow}
\usepackage{upgreek}

\def\lapp{\ifmmode\stackrel{<}{_{\sim}}\else$\stackrel{<}{_{\sim}}$\fi}
\def\gapp{\ifmmode\stackrel{>}{_{\sim}}\else$\stackrel{>}{_{\sim}}$\fi}

\newcommand{\source}{FRB~121102}
\newcommand{\frb}{\source}
\newcommand{\src}{\source}

\newcommand{\cxo}{\textit{Chandra}}
\newcommand{\cxofull}{\textit{Chandra X-ray Observatory}}
\newcommand{\chandra}{\textit{Chandra}}

\newcommand{\swift}{\textit{Swift}}
\newcommand{\fermi}{\textit{Fermi}}
\newcommand{\xmm}{\textit{XMM-Newton}}
\newcommand{\xmmnew}{\textit{XMM-Newton}}

\newcommand{\micro}{\upmu}
\newcommand{\dmunits}{\,pc\,cm$^{-3}$}
\newcommand{\nhunits}{\,cm$^{-2}$}
\newcommand{\nh}{$N_\mathrm{H}$}
\newcommand{\fluxcgs}{erg~s$^{-1}$~cm$^{-2}$}
\newcommand{\flucgs}{erg~cm$^{-2}$}
\newcommand{\lumcgs}{erg~s$^{-1}$}

\begin{document}

\title{Simultaneous X-ray, gamma-ray, and Radio Observations of the repeating Fast Radio Burst \src}

\author{P.~Scholz} 
\affil{National Research Council of Canada, Herzberg Astronomy and Astrophysics, Dominion Radio Astrophysical Observatory,
P.O. Box 248, Penticton, BC V2A 6J9, Canada; \href{mailto:paul.scholz@nrc-cnrc.gc.ca}{paul.scholz@nrc-cnrc.gc.ca}} 
\author{S.~Bogdanov}
\affil{Columbia Astrophysics Laboratory, Columbia Univ., New York, NY 10027, USA} 
\author{J.~W.~T.~Hessels}
\affil{ASTRON, the Netherlands Institute for Radio Astronomy, Postbus 2, 7990 AA Dwingeloo, The Netherlands}
\affil{Anton Pannekoek Institute for Astronomy, Univ. of Amsterdam, Science Park 904, 1098 XH Amsterdam, The Netherlands}
\author{R.~S.~Lynch}
\affil{Green Bank Observatory, PO Box 2, Green Bank, WV, 24944, USA}
\affil{Center for Gravitational Waves and Cosmology, Dept. of Physics and 
Astronomy, West Virginia Univ., White Hall, Box 6315, Morgantown, WV 
26506, USA}
\author{L.~G.~Spitler}
\affil{Max-Planck-Institut f\"ur Radioastronomie, Auf dem H\"ugel 69, 53121 Bonn, Germany}
\author{C.~G.~Bassa}
\affil{ASTRON, the Netherlands Institute for Radio Astronomy, Postbus 2, 7990 AA Dwingeloo, The Netherlands}
\author{G.~C.~Bower}
\affil{Academia Sinica Institute of Astronomy and Astrophysics, 645 N. A'ohoku Place, Hilo, HI 96720, USA}
\author{S.~Burke-Spolaor}
\affil{National Radio Astronomy Observatory, Socorro, NM 87801, USA}
\affil{Center for Gravitational Waves and Cosmology, Dept. of Physics and 
Astronomy, West Virginia Univ., White Hall, Box 6315, Morgantown, WV 
26506, USA}
\author{B.~J.~Butler}
\affil{National Radio Astronomy Observatory, Socorro, NM 87801, USA}
\author{S.~Chatterjee}
\affil{Dept. of Astronomy and Cornell Center for Astrophysics and Planetary Science, Cornell Univ., Ithaca, NY 14853, USA} 
\author{J.~M.~Cordes}
\affil{Dept. of Astronomy and Cornell Center for Astrophysics and Planetary Science, Cornell Univ., Ithaca, NY 14853, USA} 
\author{K.~Gourdji}
\affil{Anton Pannekoek Institute for Astronomy, Univ. of Amsterdam, Science Park 904, 1098 XH Amsterdam, The Netherlands}
\author{V.~M.~Kaspi}
\affil{Dept.~of Physics and McGill Space Institute, McGill Univ., Montreal, QC H3A 2T8, Canada}
\author{C.~J.~Law}
\affil{Department of Astronomy and Radio Astronomy Lab, Univ. of California, Berkeley, CA 94720, USA}
\author{B.~Marcote}
\affil{Joint Institute for VLBI ERIC, Postbus 2, 7990 AA Dwingeloo, The Netherlands}
\author{M.~A.~McLaughlin}
\affil{Center for Gravitational Waves and Cosmology, Dept. of Physics and 
Astronomy, West Virginia Univ., White Hall, Box 6315, Morgantown, WV 
26506, USA}
\author{D.~Michilli}
\affil{Anton Pannekoek Institute for Astronomy, Univ. of Amsterdam, Science Park 904, 1098 XH Amsterdam, The Netherlands}
\affil{ASTRON, the Netherlands Institute for Radio Astronomy, Postbus 2, 7990 AA Dwingeloo, The Netherlands}
\author{Z.~Paragi}
\affil{Joint Institute for VLBI ERIC, Postbus 2, 7990 AA Dwingeloo, The Netherlands}
\author{S.~M.~Ransom}
\affil{National Radio Astronomy Observatory, Charlottesville, VA 22903, USA}
\author{A.~Seymour}
\affil{Arecibo Observatory, HC3 Box 53995, Arecibo, PR 00612, USA}
\author{S.~P.~Tendulkar}
\affil{Dept.~of Physics and McGill Space Institute, McGill Univ., Montreal, QC H3A 2T8, Canada}
\author{R.~S.~Wharton}
\affil{Dept. of Astronomy and Cornell Center for Astrophysics and Planetary Science, Cornell Univ., Ithaca, NY 14853, USA} 

\begin{abstract}

We undertook coordinated campaigns with the
Green Bank, Effelsberg, and Arecibo radio telescopes during \cxofull\ and \xmmnew\
observations of the repeating fast radio burst \frb\ 
to search for simultaneous radio and X-ray bursts. 
We find 12 radio bursts from \frb\ during 70\,ks total of X-ray observations.
We detect no X-ray photons at the times of radio
bursts from \frb\ and further detect no X-ray bursts above the measured
background at any time. We place a 5$\sigma$ upper limit of $3\times10^{-11}$\,\flucgs\
on the 0.5--10\,keV fluence for X-ray bursts at the time of radio bursts for 
durations $<700$\,ms,
which corresponds to a burst energy of $4\times10^{45}$\,erg at the measured
distance of \frb.
We also place limits on the 0.5--10\,keV fluence of $5\times10^{-10}$\,\flucgs\ 
and $1\times10^{-9}$\,\flucgs\ for bursts emitted at any time during the 
\xmm\ and \cxo\ observations, respectively, 
assuming a typical X-ray burst duration of 5\,ms.
We analyze data from the {\em Fermi Gamma-ray Space Telescope} Gamma-ray Burst
Monitor and place
a 5$\sigma$ upper limit on the 10--100\,keV fluence of $4\times10^{-9}$\,\flucgs\
($5\times10^{47}$\,erg at the distance of \frb)
for gamma-ray bursts at the time of radio bursts.
We also present a deep search for a persistent X-ray source using all of the
X-ray observations taken to date and place a 5$\sigma$ upper limit on 
the 0.5--10\,keV flux of $4\times10^{-15}$\,\fluxcgs\ 
($3\times10^{41}$\,\lumcgs\ at the distance of \frb).
We discuss these non-detections in the context of the host environment of
\frb\ and of possible sources of fast radio bursts in general.

\end{abstract}


\section{Introduction}
\label{sec:intro}

Fast radio bursts (FRBs) are a recently discovered \citep{lbm+07,tsb+13} class of radio
transient that have as yet unclear physical origins. They are short (durations of
milliseconds), bright (peak flux densities $\sim0.1-10$\,Jy at 1--2\,GHz) bursts, that 
appear to be coming from outside the Galaxy based on their high dispersion
measures (DMs). Their implied distances, based on the DM excesses in comparison
to the expected line-of-sight contributions from our Galaxy \citep{cl02,ymw17}, 
suggest that they come
from cosmological redshifts \citep[i.e. $z\gapp 0.5$;][]{tsb+13}.
To date, 23 FRB sources have been discovered, 17 of which have been
found with the Parkes Telescope, one each at the Arecibo and Green Bank
Telescopes \citep{sch+14,mls+15}, three using the UTMOST array \citep{cfb+17},
and one at the Australian Square Kilometre Array Pathfinder \citep{bsm+17}.
See \citet{pbj+16} for a catalog of published 
FRBs\footnote{\url{http://www.astronomy.swin.edu.au/pulsar/frbcat/}}.

The first FRB discovered at a telescope other than Parkes was \frb\ \citep{sch+14}
at the 305-m Arecibo telescope in the PALFA Survey \citep{cfl+06,lbh+15}.
Follow-up observations of \frb\ revealed additional bursts from a location
and DM consistent with the original burst \citep{ssh+16}.
This showed that \frb\ cannot be explained by cataclysmic models 
\citep[e.g.][]{kim13,fr14}, though this may not be true of all FRBs. 
Strong arguments were also made for the 
extragalactic nature of \frb\ based on the lack of any evidence for 
any Galactic \ion{H}{2} region to provide the excess dispersing plasma \citep{ssh+16a}.

The extragalactic nature was confirmed when a direct sub-arcsecond 
localization of the repeating bursts was achieved
from Karl G. Jansky Very Large Array (VLA) observations in late 2016 \citep{clw+17}
and a host galaxy was identified.
Using optical imaging and spectroscopy with the Gemini and Keck telescopes, 
the host was found to be a faint, low-metallicity, 
star-forming, dwarf galaxy with a redshift of $z=0.193$ 
\citep[implying a luminosity distance of 972\,kpc;][]{tbc+17}.
The VLA observations also showed that the source of \frb\ is coincident with
a 0.2\,mJy persistent radio source and 
European VLBI Network (EVN) observations further showed that the persistent
source is compact to $\lapp0.2$\,mas ($\lapp0.7$\,pc, given the host distance) 
and that the bursts come from within $\lapp12$\,mas ($\lapp40$\,pc) 
of the persistent source \citep{mph+17}.
Using {\em Hubble Space Telescope} observations, \citet{bta+17} 
resolved the host galaxy and showed that the burst and persistent radio source 
is located in a bright star forming region on the outskirts of the galaxy.
Though they are co-located, and 
thus very likely share some kind of physical or evolutionary relationship, the 
persistent source and the source of radio bursts do not necessarily need to be 
one and the same.

Many models have been proposed for FRBs \citep[for a review see][]{katz16a}.
The extreme luminosities and short duration of FRBs point to coherent emission
originating from a compact object. Two classes of known phenomena that emit 
repeated radiation on those timescales are X-ray/gamma-ray bursts from 
magnetars \citep{pp13} and giant pulses from radio pulsars \citep{pc15,cw16}. 
The identification of the host galaxy of \frb\ as a low-metallicity dwarf 
\citep{tbc+17}, as well as the source's projected location in a star forming 
region \citep{bta+17},
bolsters the case for the possible magnetar nature of the 
source since these galaxies are preferentially hosts to long gamma-ray bursts 
(LGRBs) and hydrogen-poor superluminous supernovae (SLSNe-I), 
which are thought to result in 
the birth of magnetars \citep{lcb+14}. The nature of the persistent 
source in this model would be a pulsar wind nebula driven by the young 
magnetar \citep{km17} or an interaction of the supernova blast wave with 
surrounding progenitor wind bubble \citep{mbm17}.

Known Galactic magnetars produce both X-ray and gamma-ray bursts/flares and
radio pulsations on timescales of a few to hundreds of milliseconds,
similar to the durations of FRB radio bursts. 
\citet{lyu02} estimates a ratio of radio to X-ray 
energy emitted in such bursts of $10^{-4}$ based on analogies to solar flares. 
The model of \citet{lyu14} where a synchrotron maser is produced from a 
magnetized shock, predicts $10^{-5}-10^{-6}$. Given the energies of radio bursts
of FRBs, $\sim10^{39}-10^{41}$\,erg, these models predict X-ray energies
of $\sim10^{43}-10^{47}$\,erg which may be detectable by X-ray and gamma-ray 
telescopes.

Here we present a campaign of simultaneous X-ray and radio observations in
late 2016 and early 2017 with the goal of detecting or constraining any
X-ray counterparts to the radio bursts from \frb. This improves on
previous X-ray burst searches \citep[e.g.][]{ssh+16a}, which were not simultaneous
and thus sub-optimal for probing coincident X-ray bursts. In Section \ref{sec:obs}
we describe the observations performed. In Section \ref{sec:results} we
present the results of our search for radio (Arecibo, Green Bank, and 
Effelsberg Telescopes), 
X-ray (\cxo\ and \xmm) and gamma-ray (\fermi) bursts during the coordinated 
campaign as well as limits on a persistent X-ray source. We discuss the 
significance of these results in Section \ref{sec:discussion}. 

\section{Observations}
\label{sec:obs}

\begin{figure}
\includegraphics[width=\columnwidth]{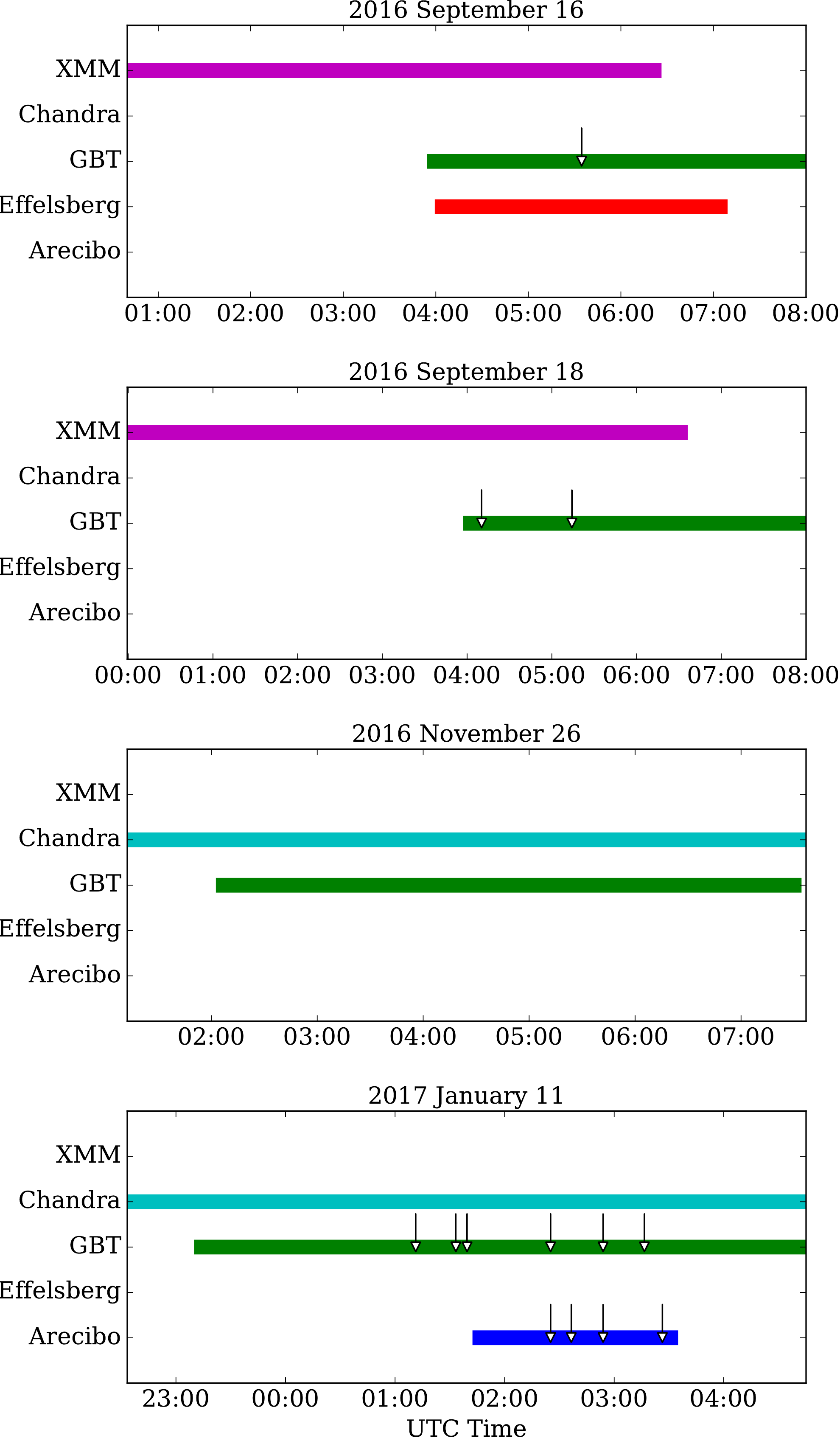}
\figcaption{
Timelines of overlapping radio and X-ray observations. Each bar represents the
time when each telescope was observing. The arrows mark the times of detected bursts. 
Note that GBT bursts 1 and 2 are only $\sim40$\,ms apart (see Figure 
\ref{fig:gbtbursts}) and so appear as
a single arrow. Note that we only present the radio burst detections that have simultaneous
X-ray coverage in this work.
\label{fig:obs}
}
\end{figure}

In 2016 September, 2016 November, and 2017 January, we undertook coordinated observations
between radio telescopes, namely the Green Bank, Arecibo, and Effelsberg telescopes,
and the \xmm\ and \cxo\ X-ray telescopes. Table \ref{tab:obs} and Figure \ref{fig:obs}
summarize the observations presented here and their overlap. In all cases, the
telescopes were pointed at the position of \frb,
R.A. = 05$^{\mathrm h}$31$^{\mathrm m}$58\fs701,
decl. = +33\arcdeg08\arcmin52\farcs55 \citep{mph+17}.
This position was also used to correct the arrival times of
the data to the solar-system barycenter (SSB).
For all data sets, the DE405 solar system ephemeris was used for
barycentric corrections.

\floattable
\begin{deluxetable*}{ccccc} 
\tabletypesize{\normalsize} 
\tablecolumns{5} 
\tablewidth{0pt} 
\tablecaption{ Summary of Joint X-ray/Radio Observations \label{tab:obs}} 
\tablehead{ 
\colhead{Telescope} & \colhead{Obs ID/}    & \colhead{Start time}    & \colhead{End time} &  \colhead{Exposure time}\\ 
\colhead{}          & \colhead{Proj. Code} & \colhead{(UTC)}  & \colhead{(UTC)}  & \colhead{(s)}}
\startdata 
\multirow{2}{*}{\xmmnew} & 0792382801 & 2016-09-16 00:39:57 & 2016-09-16 06:21:37 & 13490 \\
                      & 0792382901 & 2016-09-17 23:59:20 & 2016-09-18 06:31:00 & 15621 \\
\hline
\multirow{2}{*}{\cxo} & 19286      & 2016-11-26 01:12:24 & 2016-11-26 07:36:48 & 20810 \\
                      & 19287      & 2017-01-11 22:33:22 & 2017-01-12 04:44:43 & 19820 \\
\hline
\multirow{4}{*}{GBT}  & \multirow{2}{*}{GBT16B-391} & 2016-09-16 03:59:12 & 2016-09-16 08:00:04 & 14452 \\
                      &            & 2016-09-18 04:02:15 & 2016-09-18 08:00:04 & 14269 \\
                      & \multirow{2}{*}{CH18500414} & 2016-11-26 02:06:46 & 2016-11-26 07:30:04 & 19398 \\ 
                      &            & 2017-01-11 23:13:56 & 2017-01-12 04:45:05 & 19869 \\
\hline
Effelsberg            &            & 2016-09-16 04:04:06 & 2016-09-16 07:04:24 & 10818 \\
\hline
Arecibo               & P3054      & 2017-01-12 01:46:27 & 2017-01-12 03:30:57 & 6270 \\
\enddata 
\end{deluxetable*}

\subsection{Green Bank Telescope}

The 110-m Robert C. Byrd Green Bank Telescope (GBT) observed \frb\ on
2016 September 16, 2016 September 18, 2016 November 26, and 2017 January 11 
during periods that overlapped with either \xmm\ or \cxo\ observations 
(Table \ref{tab:obs}; Figure \ref{fig:obs}).
\frb\ was observed with the S-band receiver at a center frequency
of 2\,GHz and a bandwidth of 800\,MHz, of which about 600\,MHz is usable due 
to receiver roll-off and the masking of spectral channels containing radio 
frequency interference (RFI). 
We used the Green Bank Ultimate Pulsar Processing                                 
Instrument \citep[GUPPI;][]{drd+08} in coherent dedispersion mode where each of the spectral 
channels were corrected for dispersion in real time to DM=$557$\dmunits.
These coherently dedispersed observations therefore do not suffer from 
intra-channel DM smearing as the correction is performed before Stokes
parameters are formed.
This mode provides full Stokes parameters, 512 spectral channels 
(each 1.56\,MHz wide) and a time resolution of 10.24\,$\micro$s.

\subsection{Effelsberg Telescope}

The 100-m Effelsberg telescope performed a three hour observation of \frb\ on
2016 September 16, simultaneous with both the GBT and \xmm\ observations 
(Table \ref{tab:obs}). 
Data were recorded at 1.4 GHz with an observing configuration that was identical to 
what is used for the HTRU-N pulsar and fast transient survey. Details can be found in 
\citet{bck+13}.
\frb\ was observed with the central pixel of the 7-beam L-band receiver and data were 
recorded with the PFFTS pulsar search mode backends. The data cover a frequency 
range of 1210 to 1510 MHz with 512 frequency channels and have a time 
resolution of 54.613 $\mu$s. 
Note, unlike the GBT and Arecibo observations, these data were not coherently 
dedispersed, and so suffer from $\sim1$\,ms of intra-channel DM smearing at
the DM of \src.

\subsection{Arecibo Telescope}

The 305-m William E. Gordon Telescope at Arecibo Observatory observed \frb\
on 2017 January 12 simultaneously with GBT and \cxo\ observations 
(Table \ref{tab:obs}). We used the
single-pixel L-wide receiver with the Puerto Rican Ultimate Pulsar Processing
Instrument (PUPPI) backend. This setup provided 800\,MHz of bandwidth centered
at 1380\,MHz, of which about 600\,MHz is usable due to receiver roll-off and 
RFI excision. As with GUPPI, the data were coherently dedispersed to 
DM=$557$\dmunits\ in real time
with 512 spectral channels (each 1.56\,MHz wide) and a time 
resolution of 10.24\,$\micro$s.

\subsection{\xmmnew}

Two \xmmnew\ Director's Discretionary Time (DDT) observations 
were performed on 2016 September 16 and 18 
(Table \ref{tab:obs}). 
These DDT observations were scheduled in response to a period of 
high radio burst activity, with several bursts detected per hour
\citep{clw+17,lab+17}.
We used the EPIC/pn camera in Small Window mode (5.7-ms time resolution) 
and the EPIC/MOS cameras in Timing mode (1.75-ms time resolution). 
The Timing mode observations provide 
only one dimension of spatial information which results in a high background
(see Section \ref{sec:xraylim}).

Unfortunately, the pn-mode observations have a deadtime fraction of 29\%. 
This means that for every 5.7-ms frame of the pn observation, 
the telescope is blind to X-ray photons for 1.65\,ms. 
Though this time resolution is helpful in resolving bursts when a significant
number of counts are detected, the deadtime is detrimental when placing a limit following
a non-detection. We therefore do not use the pn-mode data below when placing fluence
limits for putative X-ray bursts.

Standard tools from the \xmm\ Science Analysis System (SAS) version 16.0 and 
HEASoft version 6.19 were used to reduce the data.
For each observation, the raw Observation Data Files (ODF) level data were 
downloaded and were pre-processed using the SAS tools {\em emproc} and {\em epproc}.
Data were filtered so that single--quadruple events with
energies between 0.1--12\,keV (pn) and 0.2--15\,keV (MOS) were retained, and
standard ``FLAG'' filtering was applied.
The light curves were then inspected for soft proton flares and none were
found.
Event arrival times were then corrected to the SSB.
For the pn, we extracted source events from an 18\arcsec\ radius (80\% encircled
energy) source region centered on the position of \frb.
For the MOS cameras, we extracted events from a 20-pixel (22\arcsec) 
wide strip centered on the position of the source.

\subsection{\cxofull}

The \cxofull\ targeted \frb\ on 2016 
November 26 (ObsID 19286) and 2017  January 11 (ObsID 19287) using the 
front-illuminated ACIS-I instrument for 20\,ks in both instances. The 
detector was operated in VFAINT mode with the entire array read out 
using a 3.2-s frame time. These observations were part of a joint 
\cxo/GBT Cycle 18 project to obtain contemporaneous data with the two 
telescopes. 

The resulting \cxo\ data sets were analyzed using 
CIAO\footnote{Chandra Interactive Analysis of Observations. 
\url{http://cxc.harvard.edu/ciao/}} version 
4.8.2 \citep{fma+06} following standard procedures recommended by 
the \cxo\ X-ray Center.
We extracted events from a 1\arcsec\ radius region (95\% encircled energy)
centered on the position of \frb\ and corrected the photon arrival times to 
the SSB
using the aforementioned source position.

\section{Analysis and Results}
\label{sec:results}

\subsection{Radio bursts}
\label{sec:radiobursts}

The radio observations from GBT, Effelsberg, and Arecibo were searched for
bursts using standard tools in the 
PRESTO\footnote{\url{http://www.cv.nrao.edu/~sransom/presto/}} \citep{ran01}
software package.
For the purposes of searching, the data were downsampled by a factor of 8
(to a time resolution of 81.92\,$\micro$s) for Arecibo and GBT observations and a 
factor of 16 (874\,$\micro$s resolution) for Effelsberg observations.
We first used {\tt rfifind} to identify frequency and time blocks contaminated 
by RFI, which were masked in the subsequent search.
The data were then dedispersed in a DM range of 507--607\dmunits\ 
with step size 0.5\dmunits\ for GBT, 535--585\dmunits\ with a step size of 
1\dmunits\ for Effelsberg, and 527--626\dmunits\ with a step size of       
1\dmunits\ for Arecibo.
Burst candidates were identified in a boxcar matched filtering search for pulse 
widths up to 20~ms and S/N~$>$~5 using {\tt single\_pulse\_search.py}.
Due to the effects of RFI, our search is only complete in the Arecibo 
observation to S/N~$\gapp$~13 and to S/N~$\gapp$~7 for Effelsberg and GBT.

Four astrophysical bursts at a DM consistent with that of \frb\ were found in the 
GBT observations on 2016 September 16 and 18 at times that overlapped with the
simultaneous \xmm\ observations (labelled, in this work, as bursts GBT 1--4). 
Radio bursts at times outside of the simultaneous X-ray coverage are not presented
here.
Five more bursts were found in the GBT observation
on 2017 January 11 that overlapped with the \cxo\ observation (bursts GBT 5--7, 9, and 10). 
Three bursts were found in the search of the 2017 January 11 Arecibo 
observation (AO 1, 2, and 4). During one of the detected GBT bursts (GBT 9) 
a coincident burst (AO 3) was found in the Arecibo observation when searching 
specifically at the time of burst GBT~9, which was otherwise missed due to RFI. 
Similarly, burst GBT~8 was found at the time of burst AO~1.
The remainder of the GBT- and Arecibo-detected bursts were not detected in the
corresponding other telescope.
To summarize, a total of 12 radio bursts were found with two co-detections 
(GBT~8/AO~1 and GBT~9/AO~3).

In the Effelsberg observation on 2016 September 16 all events with S/N~$>$~7 
can be attributed to RFI. Since these observations overlap with the times of 
GBT detected bursts, the three second windows around the arrival times of the coincident    
GBT bursts were searched manually, using a range of downsample factors,         
but no bursts were identified.
These contemporaneous non-detections, as well as those during the 2017 January 11 
Arecibo and GBT observations that did not result in co-detections, are likely due to the 
narrow-band nature of \frb\ radio bursts \citep[see][]{ssh+16,ssh+16a}.

\begin{figure*}
\includegraphics[width=\textwidth]{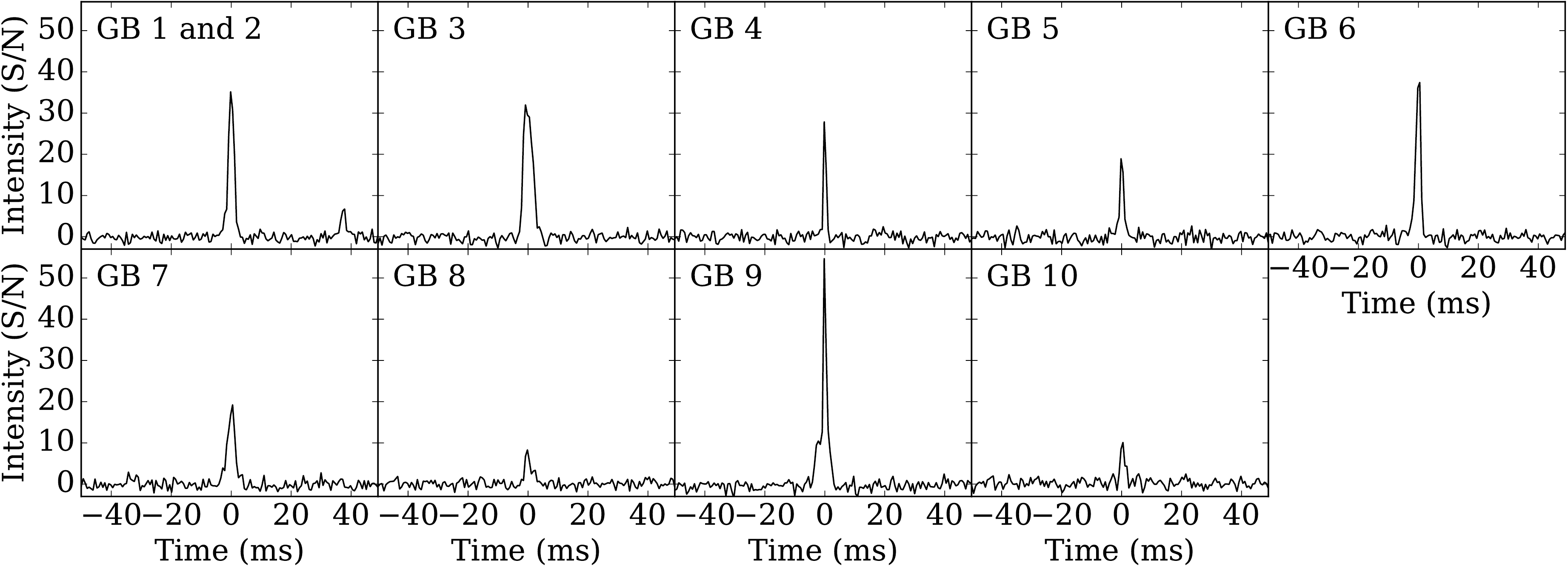}
\figcaption{
Timeseries for each GBT burst that occurred during the \xmmnew\ and \cxo\ 
observations.
Each burst has been dedispersed to 559\dmunits\ 
to be consistent with the measured average DM in \citet{ssh+16a}. 
Each timeseries has been downsampled to a time resolution of 655.36\,$\micro$s.
\label{fig:gbtbursts}
}
\end{figure*}

\begin{figure*}
\includegraphics[width=\textwidth]{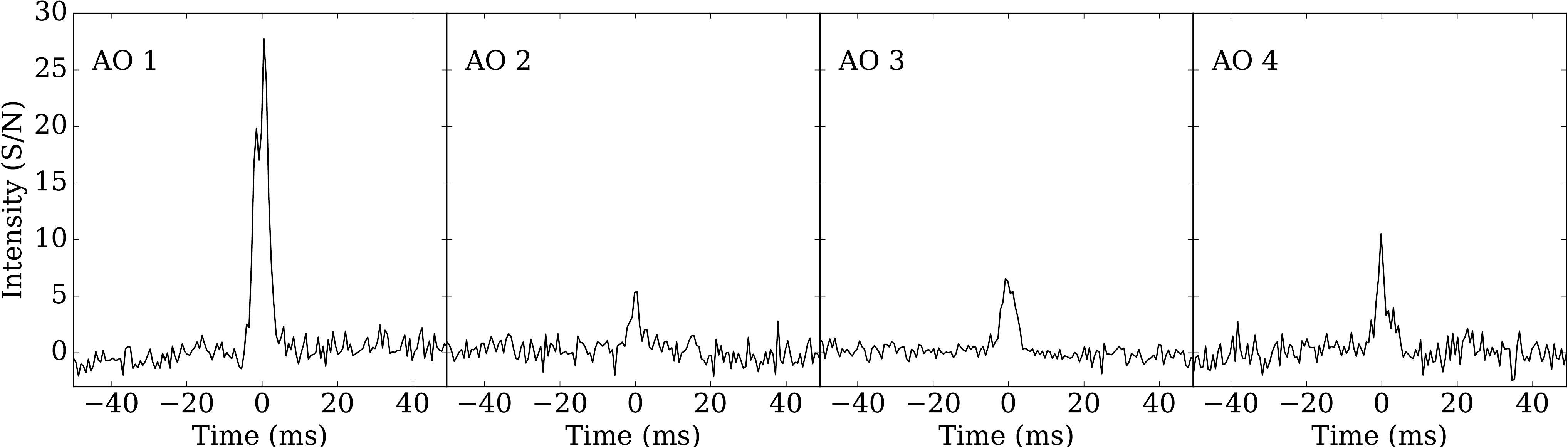}
\figcaption{
Timeseries for each Arecibo-detected burst that occurred during the \cxo\ 
observation on 2017 January 12.
Each burst has been dedispersed to 559\dmunits\ 
to be consistent with the measured average DM in \citet{ssh+16a}. 
Each timeseries has been downsampled to a time resolution of 655.36\,$\micro$s.
\label{fig:aobursts}
}
\end{figure*}

The GBT detected radio bursts are shown in Figure \ref{fig:gbtbursts} and
those detected at Arecibo are shown in Figure \ref{fig:aobursts}.
We show only the time series for each burst and defer a full spectral analysis
of the bursts to a future work.
For each burst we measured the peak flux density, fluence and burst width 
(Gaussian FWHM) using an identical procedure to \citet{ssh+16a}.
These values are shown in Table \ref{tab:bursts}.
We note that bursts GBT 1 and 2 arrived $\sim 40$\,ms apart. This is the
minimum separation reported thus far for radio bursts from \frb.
This does not necessarily imply an 
upper limit to an underlying periodicity of $<40$\,ms, as we cannot exclude 
the possibility that this is a single wide burst with multiple peaks or,
if the source is a rotating object, that multiple bursts were emitted during a 
single rotation. We defer a more detailed analysis of the arrival times to a 
future work with a much larger sample of bursts. 
The measured radio burst arrival times were corrected to the 
SSB and corrected for the dispersive delay to infinite frequency using 
DM=$559$\dmunits\ \citep{ssh+16a} (which is more accurately measured than
the DM=$557$\dmunits\ used during data collection; see Section \ref{sec:obs})
and are therefore directly comparable to the SSB-referenced arrival times of 
the X-ray photons.

\floattable
\begin{deluxetable*}{ccccccccc} 
\tabletypesize{\normalsize} 
\tablecolumns{6} 
\tablewidth{0pt} 
\tablecaption{ Detected Radio Bursts \label{tab:bursts}} 
\tablehead{ 
\colhead{Burst No.} & \colhead{Barycentric}  & \colhead{Peak Flux Density} & \colhead{Fluence} & Gaussian FWHM & X-ray Fluence Limit\tablenotemark{b} \\ 
\colhead{}          & \colhead{Arrival Time\tablenotemark{a}} & \colhead{(Jy)}      & \colhead{(Jy\,ms)}   & (ms) & (10$^{-10}$ erg\,cm$^{-2}$) } 
\startdata 
GBT 1  & 57647.232346450619 & 0.36 & 0.82 & $2.16\pm0.06$ & 2 \\
GBT 2  & 57647.232346883015 & 0.08 & 0.16 & $1.94\pm0.25$ & 2 \\
GBT 3  & 57649.173812898174 & 0.36 & 1.32 & $3.45\pm0.07$ & 2 \\
GBT 4  & 57649.218213226581 & 0.29 & 0.34 & $0.88\pm0.07$ & 2 \\
GBT 5  & 57765.049526345771 & 0.17 & 0.33 & $1.40\pm0.09$ & 5 \\
GBT 6  & 57765.064793212950 & 0.38 & 0.83 & $1.79\pm0.04$ & 5 \\
GBT 7  & 57765.069047502300 & 0.20 & 0.62 & $2.97\pm0.12$ & 5 \\
GBT 8\tablenotemark{c} & 57765.100827859293 & 0.09 & 0.18 & $2.46\pm0.28$ & 5 \\ 
GBT 9\tablenotemark{c} & 57765.120778204779 & 0.56 & 1.08 & $1.36\pm0.03$ & 5 \\
GBT 10 & 57765.136498608757 & 0.11 & 0.22 & $1.68\pm0.17$ & 5 \\
\hline
AO 1\tablenotemark{c}  & 57765.100827849608 & 0.09 & 0.37 & $4.29\pm0.11$ & 5 \\
AO 2   & 57765.108680842022 & 0.02 & 0.03 & $3.69\pm0.57$ & 5 \\
AO 3\tablenotemark{c}  & 57765.120778202479 & 0.02 & 0.05 & $4.34\pm0.44$ & 5 \\
AO 4   & 57765.143337535257 & 0.03 & 0.10 & $3.66\pm0.32$ & 5 \\
\enddata 
\tablenotetext{a}{Corrected for dispersion delay to infinite frequency using DM=559\dmunits.}
\tablenotetext{b}{5$\sigma$ confidence upper limit. See Section \ref{sec:xraylim} for details.}
\tablenotetext{c}{GBT 8/AO 1 and GBT 9/AO 3 are GBT and Arecibo co-detections (see Section \ref{sec:radiobursts}).}
\end{deluxetable*}

\subsection{Limit on X-ray burst emission}
\label{sec:xraylim}

For each detected radio burst, we searched nearby in time for X-ray photons
that could be due to X-ray bursts. In the 2016 September 16 \xmm\ observation,
the closest photon to Bursts 1 and 2 was 5.8 seconds away. The false alarm
probability for an event to arrive in such a window given the 0.5--10\,keV
background count rate of 0.01 counts\,s$^{-1}$ is 25\%.
In the 2016 September 18 \xmm\ observation no photons were closer than 0.7 seconds 
from a radio burst (background count rate 0.05 counts\,s$^{-1}$, 
false alarm probability 13\%).
In the 2017 Jan 11 \cxo\ observation, a single photon was detected at the source
position and was 893 seconds away from the closest radio burst (background 
count rate $5\times10^{-5}$ counts\,s$^{-1}$, false alarm probability 42\%).
Given the high probability that these are background events, we have no
reason to think they are related to \frb.
For each observation, the total number of X-ray counts detected within the source
extraction region of \frb\ was consistent with the background count rate. 

For both \xmm\ and \cxo\ observations we also performed a similar exercise as
above with the raw event lists (Level 1 for \cxo\ and ODF-level for \xmm) prior
to applying any standard filtering in case a bright X-ray burst was flagged as
a cosmic ray. There was no evidence for any X-ray events in excess of the 
unfiltered background count rate.

We place a limit on the number of X-ray counts from \frb\ using the Bayesian method 
of \citet{kbn91}. For a putative X-ray burst at the time of a detected radio
burst we can place an upper limit of 14.4 counts at a 5$\sigma$ confidence limit
in 0.5--10\,keV. 
This limit is independent of duration up to the time of the nearest detected 
photon (see above) because the background is negligible.
For an X-ray burst arriving outside this window at any time during the observation, 
the background rate and trials factor depend on the assumed duration. So, we 
assume a duration of 5\,ms, similar to that of the radio bursts, which leads 
to a 0.5--10\,keV count limit of 32.3 counts
during the \xmm\ observations and 33.8 counts for \cxo\ (5$\sigma$ confidence).

A count limit, $\mu_{lim}$ can be translated to a fluence limit by dividing by
a spectrally averaged effective area for the instrument, $A$, and multiplying 
by an average photon energy, $E$. 
We can also stack individual limits, under the assumption that X-ray 
bursts with similar spectra are emitted at the times of every radio burst.
In the case where zero counts are detected in each detector (i.e. \xmm/MOS and \cxo/ACIS),
and the background count rate is negligible,
the count rate limit is simply $\mu_{lim}=-\log(1-CL)$, where $CL$ is the 
desired confidence level (in our case 0.9999994 for 5$\sigma$), 
and the fluence limit takes the form,

\begin{equation}
\label{eqn:fluence}
F_{lim} = \frac{\mu_{lim}}{\sum_i \frac{N_i A_i}{E_i}}.
\end{equation}

Here the sum is over each instrument, which have each observed simultaneously
during $N_i$ radio bursts.
We read the effective area of each telescope as a function of energy from
the ancillary response files for each telescope\footnote{From
\url{http://cxc.harvard.edu/caldb/prop_plan/imaging/index.html} for \cxo/ACIS,
\url{http://xmm2.esac.esa.int/external/xmm_sw_cal/calib/epic_files.shtml} 
for \xmm/EPIC.}. 
Using these effective area curves, in Figure \ref{fig:limits} we plot the 
limiting burst energy as a function of photon energy.
Here we derive a model-independent limit where there is an equal probability
of a source photon occurring across the entire band.
The 5$\sigma$ confidence upper 
limit on the 0.5--10\,keV fluence for a single X-ray burst at the time of one 
of the detected radio bursts is $2\times10^{-10}$\,erg\,cm$^{-2}$
for observations simultaneous with \xmm\ and $5\times10^{-10}$\,erg\,cm$^{-2}$
for \cxo, or $3\times10^{46}$\,erg and $6\times10^{46}$\,erg at the luminosity
distance of \frb, respectively.
If we additionally assume that X-ray bursts of similar fluence are emitted at 
the time of every radio burst (i.e. stacked as per Equation \ref{eqn:fluence}), 
the upper limit at the time of the bursts becomes $3\times10^{-11}$\,erg\,cm$^{-2}$
($4\times10^{45}$\,erg at the source distance).
The limit for an X-ray burst arriving at any time during the X-ray observations
is $5\times10^{-10}$\,erg\,cm$^{-2}$ for \xmm\ and $1\times10^{-9}$\,erg\,cm$^{-2}$
for \cxo\ for an assumed duration of 5\,ms (which correspond to energy
limits of $6\times10^{46}$\,erg and $1\times10^{47}$\,erg).

\subsection{Limit on gamma-ray burst emission}

We also searched data from the \fermi\ Gamma-ray Burst Monitor 
\citep[GBM;][]{mlb+09} 
for gamma-ray burst counterparts during the radio bursts presented in this work.
\frb\ was only visible to GBM during the 2016 September \xmm/GBT observations,
and so our limit applies only to those four bursts. 
In an analysis similar to what has been done for previous \frb\ radio bursts
\citep{ykh+16}, we used the Time Tagged Event
GBM data in the energy range 10--100\,keV and searched for excess counts 
in 1 and 5\,ms bins in 2\,s windows centered on the arrival time of the four 
radio bursts. We find no signals that are not attributable to Poisson 
fluctuations from the background count rate at a 5$\sigma$ confidence level.
Taking into account the effective area of \fermi/GBM at 10--100\,keV, the
background count rate, and a typical photon energy of 40\,keV, we place an
upper limit of $1\times10^{-8}$\,erg\,cm$^{-2}$ for each burst and 
$4\times10^{-9}$\,erg\,cm$^{-2}$ if we assume a gamma-ray burst is emitted
at the time of each radio burst which, at the measured luminosity distance, 
corresponds to a 10--100\,keV burst energy limit of $5\times10^{47}$\,erg.

\subsection{Limit on persistent X-ray source}
\label{sec:pers}

\begin{figure}[t!]
\begin{center}
\includegraphics[width=\columnwidth]{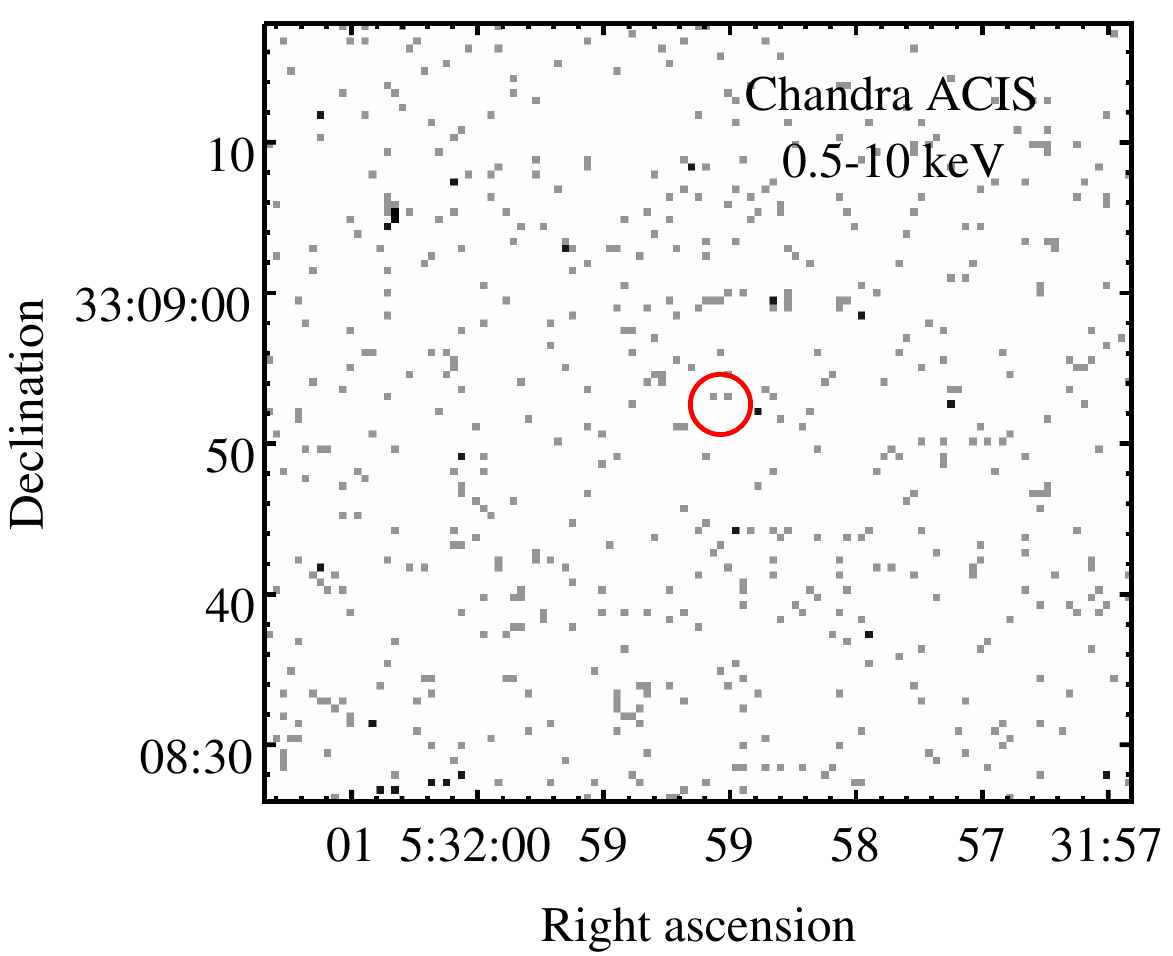}
\caption{Co-added image of all \textit{Chandra} observations of the 
\frb\ field in the 0.5--10 keV  range. The red circle of radius 
$2\arcsec$ is centered on the position from \citet{mph+17}.}
\label{fig:cxo_image}
\end{center}
\end{figure}

In order to probe more deeply for faint emission from a persistent X-ray source 
at the location of \src\ than previous limits \citep{ssh+16a,clw+17}, we produced a 
summed image of all the \cxo\ data to date.
We co-added, aspect-corrected, and exposure-corrected the two ACIS-I 
exposures from our 2016 November and 2017 January observations along with 
the ACIS-S exposure previously presented in \citet{ssh+16a} 
using the {\tt merge\_obs} script in CIAO. In the combined 80-ks exposure, 
only two events are registered within an aperture of 
radius $1\arcsec$ centered on the position of \frb\ (see Figure \ref{fig:cxo_image}), 
entirely consistent with being due to background emission. 
We measure a 0.5--10\,keV background count rate in a $40\arcsec$ radius region away from the
source to be 0.20 counts s$^{-1}$ sq.\,arcsec$^{-1}$.
Using the number of detected counts and measured background rate, we place a 
count rate limit using the Bayesian method of \citet{kbn91}, and 
translate it to a flux using the same method as in the burst case 
(Section \ref{sec:xraylim}). Assuming a 
photoelectrically absorbed power-law source spectrum with a spectral index 
of $\Gamma=2$ and a hydrogen column density of \nh$\sim1.7\times10^{22}$\,cm$^{-2}$
\citep[as in][]{clw+17}, the 5-$\sigma$ upper limit on any persistent 
0.5--10\,keV X-ray 
emission from \frb\ or the host galaxy is $4\times10^{-15}$ ergs cm$^{-2}$ s$^{-1}$. 
As the \xmm\ data presented in this work were included for the persistent
X-ray source limit in \citet{clw+17}, their value of 
$5\times10^{-15}$ ergs cm$^{-2}$ s$^{-1}$ is still valid for the \xmm\
images.

\section{Discussion}
\label{sec:discussion}

From the non-detections of X-ray photons at the times of radio bursts from
\frb\ we have placed a 0.5--10\,keV X-ray burst energy limit of 
$4\times10^{45}$\,erg assuming an X-ray burst emitted at the time
of every radio burst. If we search for bursts at any time during
X-ray observations of \frb\ we can place a 0.5--10\,keV X-ray energy limit of 
$6\times10^{46}$\,erg and $1\times10^{47}$\,erg for \xmm\ and \cxo\
observations, respectively, assuming a burst duration of 5\,ms. 
These model-independent limits, however, assume an equal probability of source
photons arriving across the entire 0.5--10\,keV band, and do not take into 
account the effects of photoelectric absorption.
These limits, therefore, can change significantly depending on the assumed 
spectral model, and come with several caveats which we will explore here.

\subsection{Effect of source models and caveats}
\label{sec:specmods}

Potential sources of X-ray bursts that accompany FRBs can 
have different underlying spectra. Here we explore the effect of different
source spectra on the fluence limits.
To generate the assumed source spectra we use {\tt XSPEC} v12.9.0n
using abundances from \citet{wam00} and photoelectric cross-sections from \citet{vfky96}
for \nh.

We initially estimate the \nh\ to the source from the DM--\nh\ relation of \citet{hnk13}.
We take the DM contribution from our Galaxy to be 188\dmunits\ 
\citep[from the NE2001 model of][]{cl02}. The DM of the host has been estimated
to be $55\lapp \mathrm{DM}_\mathrm{host} \lapp 225$\dmunits\ \citep{tbc+17}, 
so we use the average value of 140\dmunits. We assume that the IGM does not have a significant 
contribution to the \nh, as it is expected to be nearly fully ionized and thus provides
negligible X-ray absorption \citep[e.g.][]{bddl11,swt+13}.
This Galactic plus host DM of 328\dmunits\ corresponds to \nh$\sim1\times10^{22}$\,cm$^{-2}$.

However, such a determination only holds in environments similar to our Galaxy.
Photoelectric absorption and dispersion are dominated by separate components 
of the ISM, namely atomic metals and free electrons, respectively, and
their ratios could be significantly different in other environments.
To illustrate the effect of excess X-ray absorption, we also consider a case 
where the \nh\ is two orders of magnitude higher at 
$\sim1\times10^{24}$\,cm$^{-2}$. At this \nh\ nearly all of the 0.5--10\,keV
X-ray flux is absorbed. This situation may be possible in a supernova remnant 
in the first few decades following the supernova, where the ratio of atomic metals
to free electrons could be high \citep{mbm17}.

For the spectra of the bursts we assume a few fiducial models.
We take a blackbody spectrum with $kT=10$\,keV as a model similar to those
observed in magnetar hard X-ray bursts \citep[e.g.][]{lgb+12,akb+14}.
We take a cutoff power-law with index $\Gamma=0.5$ and cutoff energy of 500\,keV
as a spectrum typical of a magnetar giant flare \citep[for SGR~1806$-$20]{mca+05,pbg+05}.
Finally, we use a power-law model with index $\Gamma=2$ as an example soft
spectrum to contrast with the harder magnetar models.

\begin{figure}
\includegraphics[width=\columnwidth]{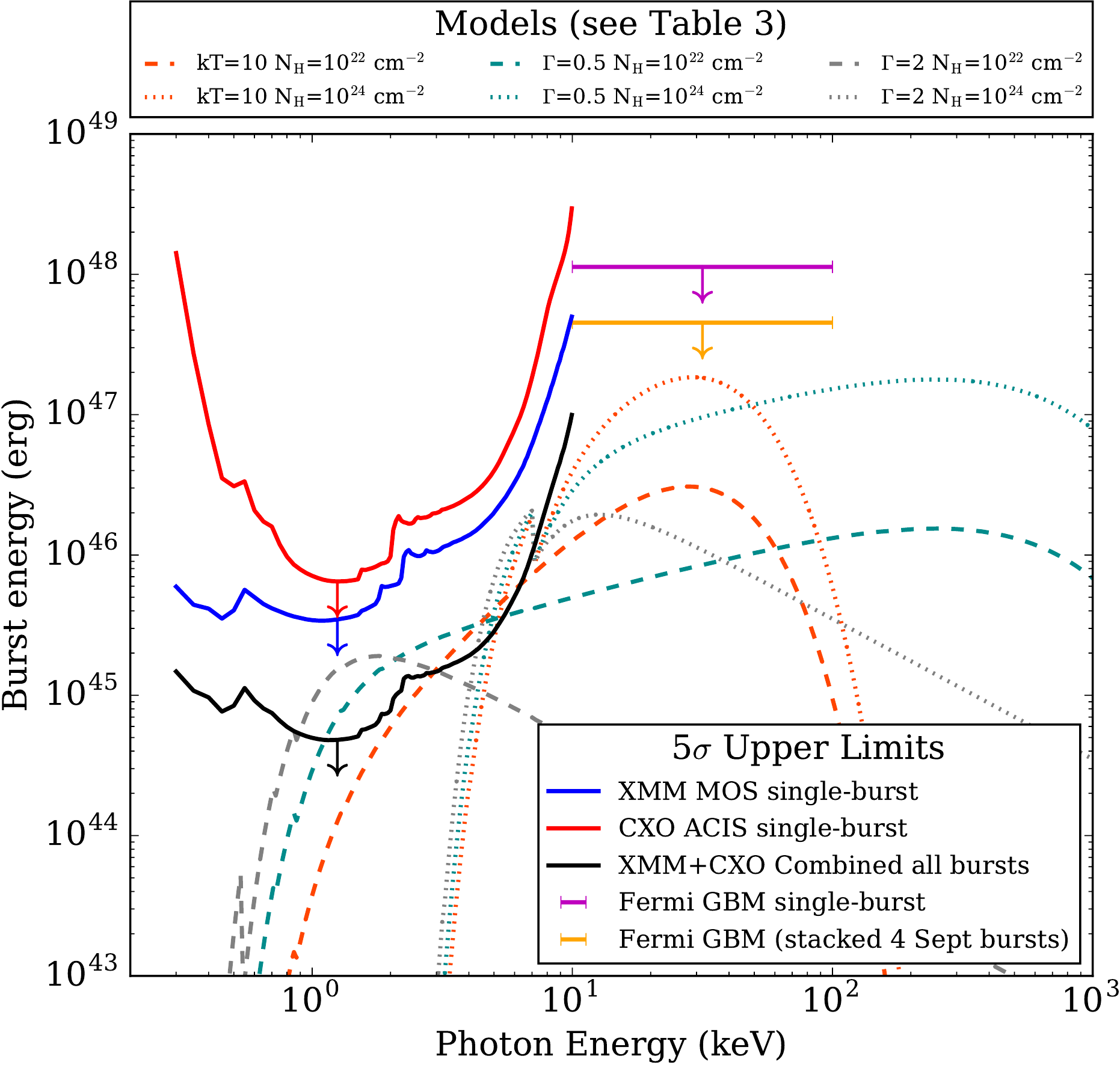}
\figcaption{
Limits on energy of X-ray bursts at the time of radio bursts from \frb.
Solid lines show the 5$\sigma$ upper limits as a function of X-ray photon
energy. 
The dashed lines show different burst spectra that are photoelectrically
absorbed by an \nh=$10^{22}$\,cm$^{-2}$ plotted at their 0.5--10\,keV fluence 
limits that result from a stacked search of the times of the radio bursts.
The dotted lines show the same spectral models but with \nh=$10^{24}$\,cm$^{-2}$ 
to show the effects of absorption. Orange lines represent a blackbody model 
with $kT=10$\,keV, cyan curves shows a cutoff power-law model with $\Gamma=0.5$ 
and $E_\mathrm{cut}=500$\,keV, and the grey curves show a soft power-law with
$\Gamma=2$ in order to illustrate the effect of different spectral models.
\label{fig:limits}
}
\end{figure}

In Table \ref{tab:lims} we give the fluence limits for each of these models
and the implied limit on their unabsorbed emitted energy both in the 0.5--10\,keV
band and extrapolated to the 10\,keV--1\,MeV gamma-ray band. Note that the
energy limits are highly dependent on the amount of absorption and the gamma-ray
energy is heavily dependent on the assumed spectrum, as it is extrapolated
outside of the 0.5--10\,keV band. In Figure \ref{fig:limits} we plot each model 
normalized to its fluence upper limit.

It is clear that assumptions on the underlying spectral model change the 
implication for the X-ray--gamma-ray luminosity. If the X-ray absorption is 
increased, the energy limits for the models in Table \ref{tab:lims} 
increase by 1--2 orders of magnitude. Furthermore, a burst that primarily emits
energy outside of the 0.5--10\,keV soft X-ray band, similar to the magnetar
burst and flare models in Table \ref{tab:lims}, can be much more luminous
than in a soft model and be undetectable by \cxo\ and \xmm.

\floattable
\begin{deluxetable}{ccccccccc} 
\tabletypesize{\normalsize} 
\tablecolumns{5} 
\tablewidth{0pt} 
\tablecaption{ Burst limits for different X-ray spectral models \label{tab:lims}}
\tablehead{ 
\colhead{Model} & \colhead{\nh}  & \colhead{kT/$\Gamma$}   & \colhead{Absorbed 0.5--10 keV}  & \colhead{Unabsorbed 0.5--10 keV} & Extrapolated 10 keV--1 MeV \\ 
\colhead{}      & \colhead{(\nhunits)} & \colhead{(keV/-)} & \colhead{Fluence Limit}         & \colhead{Energy Limit\tablenotemark{a} }           & Energy Limit\tablenotemark{a} \\    
\colhead{}      & \colhead{}     & \colhead{}              & \colhead{($10^{-11}$\,\flucgs)} & \colhead{($10^{45}$\,erg)}  & \colhead{($10^{47}$\,erg)} } 
\startdata 
Blackbody & $10^{22}$ & 10  & $5$  & 6   & $2$ \\
Blackbody & $10^{24}$ & 10  & $13$ & 110 & $30$ \\
Cutoff PL & $10^{22}$ & 0.5 & $3$  & 4   & $13$ \\
Cutoff PL & $10^{24}$ & 0.5 & $11$ & 120 & $400$ \\
Soft PL   & $10^{22}$ & 2   & 1.3  & 3   & 0.04 \\
Soft PL   & $10^{24}$ & 2   & 8    & 300 & 40 \\
\enddata 
\tablenotetext{a}{Assuming the measured luminosity distance to \frb, 972\,Mpc \citep{tbc+17}.}
\tablecomments{5$\sigma$ confidence upper limits. See Section \ref{sec:specmods} for details.}
\end{deluxetable}

When searching for X-rays at the time of radio bursts and placing an associated
limit, we are assuming 
that an X-ray burst is emitted at the same time
as each radio burst and that periods of radio burst activity are correlated
with X-ray activity.
However, if the episodic detection of radio bursts is not intrinsic but due
to amplification of the intrinsic emission from lensing by the intervening
medium \citep[e.g.][]{cwh+17}, the detection
of radio bursts and X-ray bursts may not be strongly correlated. 
If the radio bursts are externally amplified, their intrinsic energies could 
be lower by $\lapp10^2$ than what is implied from their 
detection \citep{cwh+17}. So, if the
intrinsic source of FRBs also produces X-ray emission with a fluence ratio 
$F_R/F_X$, the expected X-ray emission could be up to two orders of magnitude
lower if the radio burst is extrinsically amplified.

\subsection{Comparison to previous limits}

For all of the known FRBs at the time, \citet{tkp16} placed a limit on the 
fluence ratio defined as $\eta=F_\mathrm{1.4\,GHz} / F_\gamma$,
where $F_\mathrm{1.4\,GHz}$ is the radio fluence at a frequency of 1.4\,GHz and
$F_\gamma$ is the gamma-ray fluence.
The most constraining limit is for FRB~010724 with 
$\eta>8\times10^8$\,Jy\,ms\,erg$^{-1}$\,cm$^2$.
{If we assume our fiducial giant flare model (i.e. cut-off powerlaw with
$\Gamma=0.5$ and cutoff energy of 500\,keV) with \nh=$10^{22}$\nhunits}, 
our gamma-ray fluence limit is $1\times10^{-10}$\,\flucgs\ for a typical 
photon energy of 20\,keV \citep[as in][]{tkp16}.
If we further assume that the 1.4\,GHz fluence is approximately the same as the 2\,GHz 
fluence, we can place a lower limit on the ratio
of radio to gamma-ray fluence of $\eta>6\times10^{9}$\,Jy\,ms\,erg$^{-1}$\,cm$^2$ 
for \frb. We can also compare to our \fermi/GBM limits which, though less
constraining, do not rely on an extrapolation from soft X-rays into gamma-ray
wavelengths. Using this limit the corresponding fluence ratio limit is
$\eta>2\times10^{8}$\,Jy\,ms\,erg$^{-1}$\,cm$^2$. 

A gamma-ray burst counterpart to FRB~131104 with energy of $5\times10^{51}$\,erg
has been claimed by \citet{dfm+16}, though it has been contested by \citet{sr17}. 
The implied radio to gamma-ray fluence ratio from the claimed detection
is $\eta=6\times10^{5}$\,Jy\,ms\,erg$^{-1}$\,cm$^2$.
\cite{dfm+16} also searched for \swift/BAT sub-threshold events at 
any time \swift/BAT was pointed towards the source for 16 FRBs,
including \frb, and concluded that there is no evidence for repeated 
gamma-ray emission from those FRBs above \swift/BAT sensitivities.
We can nevertheless compare our limits to an event similar to that claimed
for FRB~131104.
Our limits clearly rule out an event of that magnitude at any time during
\xmm\ or \chandra\ observations of \frb. Further, such an event is clearly ruled
out from the \fermi/GBM limits, both at the time of the bursts and at any time
while \frb\ is visible to GBM and actively emitting radio bursts. 

\subsection{Models of FRBs}

Models of FRBs from magnetars predict a small ratio between the fluence of
radio and high-energy emission. Since the radio is a small fraction of the total
emitted energy in these models, high-energy emission may be detectable.
Based on analogies to solar flares, \citet{lyu02} estimates a ratio of $10^{-4}$. 
\citet{lyu14} predicts $10^{-5}-10^{-6}$ based on a model of a
synchrotron maser produced from a magnetized shock during magnetar activity.
Our X-ray limit implies a radio-to-X-ray ratio
of $>10^{-6}-10^{-8}$, depending on the spectral model and the absorbing column, 
which is close to the range where we can begin constraining these models.

We can also compare our limits to the most luminous magnetar giant flare emitted 
in our Galaxy, that of SGR~1806$-$20. 
The 2004 giant flare had a spectrum similar to that of our canonical giant flare in 
Table~\ref{tab:lims}, a gamma-ray luminosity of $\sim10^{47}$\,\lumcgs, 
and a duration of $\sim100$\,ms \citep{mca+05,pbg+05}. 
For the \nh$=1\times10^{22}$\,cm$^{-2}$ case, this corresponds to a 0.5--10\,keV 
fluence of $\sim5\times10^{-12}$\,\flucgs\ at the luminosity distance of \frb. 
This is approximately an order of magnitude lower than our corresponding 
extrapolated limit.

\subsection{The nature of the persistent source}

\citet{clw+17} showed that
\frb\ is associated with a $\sim0.2$\,mJy persistent radio source and \citet{mph+17}
further showed, using European VLBI Network observations, that the persistent 
source is compact to a projected size of 0.7\,pc and consistent with being 
coincident with the source of the \frb\ bursts. 
Two possible scenarios are considered in 
\citet{mph+17} for the persistent source: an extragalactic neutron star embedded
in a supernova remnant (SNR), perhaps producing a pulsar wind nebula (PWN). 
Alternatively, an AGN origin is considered. 
Here we explore how the limit on X-ray emission from a persistent source
at the location of \frb\ informs these scenarios.

At the luminosity distance of \frb, 972\,Mpc \citep{tbc+17}, no nebula similar to those
in our Galaxy would be visible by several orders of magnitude (e.g. the Crab
nebula would have a 0.5--10\,keV X-ray flux of 
$\sim1\times10^{-19}$\,\fluxcgs, well below the sensitivities of current X-ray
detectors). 
However, given the possibility that the radio
bursts of \frb\ originate from a young neutron star \citep[e.g.][]{cw16,lbp16,katz16},
a nebula resulting from either the supernova that produced the neutron star
or a wind driven by either the rotational (PWN) or 
magnetic (magnetar wind nebula) energy of the young neutron star is an attractive model
for the persistent radio source \citep{mbm17}. 

Given the existence of a luminous persistent radio source at such a distance,
one might also expect an exceptionally luminous X-ray source. 
Taking the Crab nebula and scaling
its X-ray flux by the ratio between its radio luminosity and the radio luminosity
of the persistent counterpart to \frb\ (a factor of $4\times10^5$), we arrive at a 
0.5--10\,keV X-ray flux of $\sim5\times10^{-14}$\,\fluxcgs. This is over an order
of magnitude brighter than the 5$\sigma$ limit that we placed in Section 
\ref{sec:pers}. We can therefore confidently rule out a scaled version of the
Crab nebula. However, such a nebula, powered by a young ($< 100$\,yr)
pulsar or magnetar, does not have an analogue in our Galaxy, and 
may therefore have properties different from that of the Crab. Furthermore, in 
the case that the nebula is a bright X-ray emitter, the soft X-rays may be
absorbed by the supernova ejecta \citep{mbm17}. For example, assuming a
Crab-like X-ray spectrum, our hypothetical `scaled-Crab' nebula would have an 
absorbed 0.5--10\,keV X-ray flux below our 5$\sigma$ limit for
\nh $\gapp5\times10^{23}$\,\nhunits.

The persistent source coincident with \frb\ could also plausibly be a 
massive black hole, resembling the properties observed in low-luminosity 
active galactic nuclei (LLAGNs). It was shown by \citet{mph+17} that the 
observed persistent radio emission cannot be explained by either a 
stellar-mass black hole (such as in X-ray binaries) or an intermediate-mass 
black hole. On the other hand, the mass of a super-massive black hole is 
constrained by the stellar mass of the dwarf galaxy \citep{tbc+17}. 
We would thus expect a black hole mass in the range 
$\sim 10^{5}-10^{7}\ \mathrm{M_{\odot}}$.

Considering our upper-limit on the X-ray emission (which implies a luminosity of
$\lesssim 10^{41}\ \mathrm{erg\ s^{-1}}$) and the radio flux density at 5 GHz 
\citep{clw+17,mph+17} we infer a ratio between the radio and X-ray luminosities 
\citep{tw03} of $\log R_{\rm X} \gtrsim -2.7$, consistent with the ratio observed 
in radio-loud AGNs \citep{ho08}. This ratio is also consistent with the values of 
$R_{\rm X} \sim -2$ observed in radio-loud LLAGNs \citep{psd+12} and so a massive black 
hole resembling a radio-loud LLAGNs but scaled down in mass remains a plausible 
model for the persistent radio source. 

\section{Conclusion}

Here we have placed the deepest limits to date on soft (0.5--10\,keV) X-ray
emission emitted during FRBs as well as from persistent X-ray emission from the 
location of an FRB source. These limits rule out extreme scenarios but allow many
reasonable models. 
Our limits on the 0.5-10 keV burst energy at the time of radio bursts range from 
$10^{45}$ to $10^{47}$\,erg depending on the underlying model and level of 
X-ray absorption. We can confidently rule out events with GRB-like energies \
($\gapp10^{49}$\,erg). Our limits, however, are about an order of magnitude 
higher than the brightest observed Galactic giant flare, so we do not rule
out that model.
However, X-ray bursts
from possible FRB sources, like magnetars, may emit the majority of their
flux at photon energies higher than 10\,keV. Pointed hard X-ray/soft gamma-ray
observations with telescopes such as {\em NuSTAR} will therefore be interesting.

Our limit on the persistent luminosity of an X-ray source at the location of 
the \frb\ source is $3\times10^{41}$\,\lumcgs. We showed that if we assume that
the persistent radio source at the location of \frb\ has a Crab Nebula-like spectrum,
it should have been detectable in our X-ray observations. However, such a nebula
could be undetectable if there is a high amount of X-ray absorption, or if the 
nebular spectrum has a higher radio-X-ray luminosity ratio than that of the Crab Nebula. 
We also show that the radio-X-ray luminosity ratio limit is consistent with known 
radio-loud LLAGNs.

More fundamentally, at $\sim$Gpc distances, the fluxes we expect for shorter 
wavelength (i.e. optical--X-ray--gamma-ray) counterparts are unreachable at millisecond 
timescales if the energy emitted at those wavelengths is comparable to the
emitted radio energy of FRBs. Therefore, if an event that 
produces an FRB emits a large fraction of its energy at radio wavelengths we would
not expect multi-wavelength burst counterparts to be detectable.
However, we must place the most stringent limits possible in case the converse,
that the radio emission is a small fraction of the total emitted energy, is true.
Such limits therefore inform possible models of FRBs.
In the future, with instruments that promise to detect large numbers of FRBs
(e.g. CHIME, UTMOST, DSA-10, ALERT) we may accumulate a sample of relatively
nearby (\lapp100\,Mpc), bright, repeating FRBs which would be more likely
to have detectable high-energy burst counterparts.

\acknowledgements
P.S. is a Covington Fellow at DRAO.

S.B. and S.C. acknowledge support from NASA Chandra grants GO7-18059A and GO7-18059B 
issued by the Chandra X-ray Observatory Center, which is operated by the 
Smithsonian Astrophysical Observatory for and on behalf of NASA under contract 
NAS8-03060.

J.W.T.H. acknowledges funding from an NWO Vidi fellowship. 
J.W.T.H. and C.G.B. acknowledge support from the European Research Council
under the European Union's Seventh Framework Programme (FP/2007-2013) 
/ ERC Grant Agreement nr.\ 337062 (DRAGNET; PI Hessels).

L.G.S. gratefully acknowledges support from the ERC Starting Grant
BEACON under contract no. 279702 and the Max Planck Society.

V.M.K. holds the Lorne Trottier and a Canada Research Chair and receives
support from an NSERC Discovery Grant and Herzberg Prize, from an R. Howard 
Webster Foundation Fellowship from the Canadian Institute for Advanced Research
(CIFAR), and from the FRQNT Centre de Recherche en
Astrophysique du Quebec.

C.J.L. is supported by the NSF under Grant No. 1611606.

B.M. acknowledges support from the Spanish Ministerio de Econom\'ia y 
Competitividad (MINECO) under grants AYA2016-76012-C3-1-P and MDM-2014-0369 of 
ICCUB (Unidad de Excelencia ``Mar\'ia de Maeztu'').

The Green Bank Observatory is a facility of the National Science Foundation 
operated under cooperative agreement by Associated Universities, Inc.

The Arecibo Observatory is operated by SRI International under a cooperative 
agreement with the National Science Foundation (AST-1100968), and in alliance 
with Ana G. M\'endez-Universidad Metropolitana, and the Universities Space 
Research Association. 

Based on observations with the 100-m telescope of the MPIfR
(Max-Planck-Institut f\"ur Radioastronomie) at Effelsberg.

Based on observations obtained with XMM-Newton, an ESA science mission with 
instruments and contributions directly funded by ESA Member States and NASA.

This research has made use of data obtained from the Chandra Data Archive, 
and software provided by the Chandra X-ray Center (CXC) in the application 
package CIAO.

\facilities{GBT (GUPPI), Arecibo (PUPPI), Effelsberg (PFFTS), 
XMM (EPIC/pn, EPIC/MOS), CXO (ACIS-I), Fermi (GBM)}

\software{PRESTO, XMM SAS v16.0, HEASoft v6.19, CIAO v4.8.2}

\bibliographystyle{aasjournal}
\bibliography{journals_apj,myrefs,modrefs,psrrefs,crossrefs}

\begin{thebibliography}{}
\expandafter\ifx\csname natexlab\endcsname\relax\def\natexlab#1{#1}\fi

\bibitem[{{An} {et~al.}(2014){An}, {Kaspi}, {Beloborodov}, {Kouveliotou},
  {Archibald}, {Boggs}, {Christensen}, {Craig}, {Gotthelf}, {Grefenstette},
  {Hailey}, {Harrison}, {Madsen}, {Mori}, {Stern}, \& {Zhang}}]{akb+14}
{An}, H., {Kaspi}, V.~M., {Beloborodov}, A.~M., {et~al.} 2014, ApJ, 790, 60

\bibitem[{{Bannister} {et~al.}(2017){Bannister}, {Shannon}, {Macquart},
  {Flynn}, {Edwards}, {O'Neill}, {Os{\l}owski}, {Bailes}, {Zackay}, {Clarke},
  {D'Addario}, {Dodson}, {Hall}, {Jameson}, {Jones}, {Navarro}, {Trinh},
  {Allison}, {Anderson}, {Bell}, {Chippendale}, {Collier}, {Heald}, {Heywood},
  {Hotan}, {Lee-Waddell}, {Madrid}, {Marvil}, {McConnell}, {Popping},
  {Voronkov}, {Whiting}, {Allen}, {Bock}, {Brodrick}, {Cooray}, {DeBoer},
  {Diamond}, {Ekers}, {Gough}, {Hampson}, {Harvey-Smith}, {Hay}, {Hayman},
  {Jackson}, {Johnston}, {Koribalski}, {McClure-Griffiths}, {Mirtschin}, {Ng},
  {Norris}, {Pearce}, {Phillips}, {Roxby}, {Troup}, \& {Westmeier}}]{bsm+17}
{Bannister}, K.~W., {Shannon}, R.~M., {Macquart}, J.-P., {et~al.} 2017, ApJ,
  841, L12

\bibitem[{{Barr} {et~al.}(2013){Barr}, {Champion}, {Kramer}, {Eatough},
  {Freire}, {Karuppusamy}, {Lee}, {Verbiest}, {Bassa}, {Lyne}, {Stappers},
  {Lorimer}, \& {Klein}}]{bck+13}
{Barr}, E.~D., {Champion}, D.~J., {Kramer}, M., {et~al.} 2013, MNRAS, 435, 2234

\bibitem[{{Bassa} {et~al.}(2017){Bassa}, {Tendulkar}, {Adams}, {Maddox},
  {Bogdanov}, {Bower}, {Burke-Spolaor}, {Butler}, {Chatterjee}, {Cordes},
  {Hessels}, {Kaspi}, {Law}, {Marcote}, {Paragi}, {Ransom}, {Scholz},
  {Spitler}, \& {van Langevelde}}]{bta+17}
{Bassa}, C.~G., {Tendulkar}, S.~P., {Adams}, E.~A.~K., {et~al.} 2017, ApJ, 843,
  L8

\bibitem[{{Behar} {et~al.}(2011){Behar}, {Dado}, {Dar}, \& {Laor}}]{bddl11}
{Behar}, E., {Dado}, S., {Dar}, A., \& {Laor}, A. 2011, ApJ, 734, 26

\bibitem[{{Caleb} {et~al.}(2017){Caleb}, {Flynn}, {Bailes}, {Barr}, {Bateman},
  {Bhandari}, {Campbell-Wilson}, {Farah}, {Green}, {Hunstead}, {Jameson},
  {Jankowski}, {Keane}, {Parthasarathy}, {Ravi}, {Rosado}, {van Straten}, \&
  {Venkatraman Krishnan}}]{cfb+17}
{Caleb}, M., {Flynn}, C., {Bailes}, M., {et~al.} 2017, MNRAS, 468, 3746

\bibitem[{{Chatterjee} {et~al.}(2017){Chatterjee}, {Law}, {Wharton},
  {Burke-Spolaor}, {Hessels}, {Bower}, {Cordes}, {Tendulkar}, {Bassa},
  {Demorest}, {Butler}, {Seymour}, {Scholz}, {Abruzzo}, {Bogdanov}, {Kaspi},
  {Keimpema}, {Lazio}, {Marcote}, {McLaughlin}, {Paragi}, {Ransom}, {Rupen},
  {Spitler}, \& {van Langevelde}}]{clw+17}
{Chatterjee}, S., {Law}, C.~J., {Wharton}, R.~S., {et~al.} 2017, Nature, 541,
  58

\bibitem[{{Cordes} \& {Lazio}(2002)}]{cl02}
{Cordes}, J.~M., \& {Lazio}, T.~J.~W. 2002, ArXiv:astro-ph/0207156,
  arXiv:astro-ph/0207156

\bibitem[{{Cordes} \& {Wasserman}(2016)}]{cw16}
{Cordes}, J.~M., \& {Wasserman}, I. 2016, MNRAS, 457, 232

\bibitem[{{Cordes} {et~al.}(2017){Cordes}, {Wasserman}, {Hessels}, {Lazio},
  {Chatterjee}, \& {Wharton}}]{cwh+17}
{Cordes}, J.~M., {Wasserman}, I., {Hessels}, J.~W.~T., {et~al.} 2017, ApJ, 842,
  35

\bibitem[{{Cordes} {et~al.}(2006){Cordes}, {Freire}, {Lorimer}, {Camilo},
  {Champion}, {Nice}, {Ramachandran}, {Hessels}, {Vlemmings}, {van Leeuwen},
  {Ransom}, {Bhat}, {Arzoumanian}, {McLaughlin}, {Kaspi}, {Kasian}, {Deneva},
  {Reid}, {Chatterjee}, {Han}, {Backer}, {Stairs}, {Deshpande}, \&
  {Faucher-Gigu{\`e}re}}]{cfl+06}
{Cordes}, J.~M., {Freire}, P.~C.~C., {Lorimer}, D.~R., {et~al.} 2006, ApJ, 637,
  446

\bibitem[{{DeLaunay} {et~al.}(2016){DeLaunay}, {Fox}, {Murase},
  {M{\'e}sz{\'a}ros}, {Keivani}, {Messick}, {Mostaf{\'a}}, {Oikonomou}, {Te{\v
  s}i{\'c}}, \& {Turley}}]{dfm+16}
{DeLaunay}, J.~J., {Fox}, D.~B., {Murase}, K., {et~al.} 2016, ApJ, 832, L1

\bibitem[{{DuPlain} {et~al.}(2008){DuPlain}, {Ransom}, {Demorest}, {Brandt},
  {Ford}, \& {Shelton}}]{drd+08}
{DuPlain}, R., {Ransom}, S., {Demorest}, P., {et~al.} 2008, in Society of
  Photo-Optical Instrumentation Engineers (SPIE) Conference Series, Vol. 7019,
  Society of Photo-Optical Instrumentation Engineers (SPIE) Conference Series,
  70191D

\bibitem[{{Falcke} \& {Rezzolla}(2014)}]{fr14}
{Falcke}, H., \& {Rezzolla}, L. 2014, A\&A, 562, A137

\bibitem[{{Fruscione} {et~al.}(2006){Fruscione}, {McDowell}, {Allen},
  {Brickhouse}, {Burke}, {Davis}, {Durham}, {Elvis}, {Galle}, {Harris},
  {Huenemoerder}, {Houck}, {Ishibashi}, {Karovska}, {Nicastro}, {Noble},
  {Nowak}, {Primini}, {Siemiginowska}, {Smith}, \& {Wise}}]{fma+06}
{Fruscione}, A., {McDowell}, J.~C., {Allen}, G.~E., {et~al.} 2006, in Society
  of Photo-Optical Instrumentation Engineers (SPIE) Conference Series, Vol.
  6270, Society of Photo-Optical Instrumentation Engineers (SPIE) Conference
  Series

\bibitem[{{He} {et~al.}(2013){He}, {Ng}, \& {Kaspi}}]{hnk13}
{He}, C., {Ng}, C.-Y., \& {Kaspi}, V.~M. 2013, ApJ, 768, 64

\bibitem[{{Ho}(2008)}]{ho08}
{Ho}, L.~C. 2008, Ann. Rev. Astr. Ap., 46, 475

\bibitem[{{Kashiyama} {et~al.}(2013){Kashiyama}, {Ioka}, \&
  {M{\'e}sz{\'a}ros}}]{kim13}
{Kashiyama}, K., {Ioka}, K., \& {M{\'e}sz{\'a}ros}, P. 2013, ApJ, 776, L39

\bibitem[{{Kashiyama} \& {Murase}(2017)}]{km17}
{Kashiyama}, K., \& {Murase}, K. 2017, ApJ, 839, L3

\bibitem[{{Katz}(2016{\natexlab{a}})}]{katz16a}
{Katz}, J.~I. 2016{\natexlab{a}}, Modern Physics Letters A, 31, 1630013

\bibitem[{{Katz}(2016{\natexlab{b}})}]{katz16}
---. 2016{\natexlab{b}}, ApJ, 826, 226

\bibitem[{{Kraft} {et~al.}(1991){Kraft}, {Burrows}, \& {Nousek}}]{kbn91}
{Kraft}, R.~P., {Burrows}, D.~N., \& {Nousek}, J.~A. 1991, ApJ, 374, 344

\bibitem[{{Law} {et~al.}(2017){Law}, {Abruzzo}, {Bassa}, {Bower},
  {Burke-Spolaor}, {Butler}, {Cantwell}, {Carey}, {Chatterjee}, {Cordes},
  {Demorest}, {Dowell}, {Fender}, {Gourdji}, {Grainge}, {Hessels}, {Hickish},
  {Kaspi}, {Lazio}, {McLaughlin}, {Michilli}, {Mooley}, {Perrott}, {Ransom},
  {Razavi-Ghods}, {Rupen}, {Scaife}, {Scott}, {Scholz}, {Seymour}, {Spitler},
  {Stovall}, {Tendulkar}, {Titterington}, {Wharton}, \& {Williams}}]{lab+17}
{Law}, C.~J., {Abruzzo}, M.~W., {Bassa}, C.~G., {et~al.} 2017, ApJ,
  arXiv:1705.07553, submitted

\bibitem[{{Lazarus} {et~al.}(2015){Lazarus}, {Brazier}, {Hessels},
  {Karako-Argaman}, {Kaspi}, {Lynch}, {Madsen}, {Patel}, {Ransom}, {Scholz},
  {Swiggum}, {Zhu}, {Allen}, {Bogdanov}, {Camilo}, {Cardoso}, {Chatterjee},
  {Cordes}, {Crawford}, {Deneva}, {Ferdman}, {Freire}, {Jenet}, {Knispel},
  {Lee}, {van Leeuwen}, {Lorimer}, {Lyne}, {McLaughlin}, {Siemens}, {Spitler},
  {Stairs}, {Stovall}, \& {Venkataraman}}]{lbh+15}
{Lazarus}, P., {Brazier}, A., {Hessels}, J.~W.~T., {et~al.} 2015, ApJ, 812, 81

\bibitem[{{Lin} {et~al.}(2012){Lin}, {G{\"o}{\v g}{\"u}{\c s}}, {Baring},
  {Granot}, {Kouveliotou}, {Kaneko}, {van der Horst}, {Gruber}, {von Kienlin},
  {Younes}, {Watts}, \& {Gehrels}}]{lgb+12}
{Lin}, L., {G{\"o}{\v g}{\"u}{\c s}}, E., {Baring}, M.~G., {et~al.} 2012, ApJ,
  756, 54

\bibitem[{{Lorimer} {et~al.}(2007){Lorimer}, {Bailes}, {McLaughlin},
  {Narkevic}, \& {Crawford}}]{lbm+07}
{Lorimer}, D.~R., {Bailes}, M., {McLaughlin}, M.~A., {Narkevic}, D.~J., \&
  {Crawford}, F. 2007, Science, 318, 777

\bibitem[{{Lunnan} {et~al.}(2014){Lunnan}, {Chornock}, {Berger}, {Laskar},
  {Fong}, {Rest}, {Sanders}, {Challis}, {Drout}, {Foley}, {Huber}, {Kirshner},
  {Leibler}, {Marion}, {McCrum}, {Milisavljevic}, {Narayan}, {Scolnic},
  {Smartt}, {Smith}, {Soderberg}, {Tonry}, {Burgett}, {Chambers}, {Flewelling},
  {Hodapp}, {Kaiser}, {Magnier}, {Price}, \& {Wainscoat}}]{lcb+14}
{Lunnan}, R., {Chornock}, R., {Berger}, E., {et~al.} 2014, ApJ, 787, 138

\bibitem[{{Lyubarsky}(2014)}]{lyu14}
{Lyubarsky}, Y. 2014, MNRAS, 442, L9

\bibitem[{Lyutikov(2002)}]{lyu02}
Lyutikov, M. 2002, ApJ, 580, L65

\bibitem[{{Lyutikov} {et~al.}(2016){Lyutikov}, {Burzawa}, \& {Popov}}]{lbp16}
{Lyutikov}, M., {Burzawa}, L., \& {Popov}, S.~B. 2016, MNRAS, 462, 941

\bibitem[{{Marcote} {et~al.}(2017){Marcote}, {Paragi}, {Hessels}, {Keimpema},
  {van Langevelde}, {Huang}, {Bassa}, {Bogdanov}, {Bower}, {Burke-Spolaor},
  {Butler}, {Campbell}, {Chatterjee}, {Cordes}, {Demorest}, {Garrett}, {Ghosh},
  {Kaspi}, {Law}, {Lazio}, {McLaughlin}, {Ransom}, {Salter}, {Scholz},
  {Seymour}, {Siemion}, {Spitler}, {Tendulkar}, \& {Wharton}}]{mph+17}
{Marcote}, B., {Paragi}, Z., {Hessels}, J.~W.~T., {et~al.} 2017, ApJ, 834, L8

\bibitem[{{Masui} {et~al.}(2015){Masui}, {Lin}, {Sievers}, {Anderson}, {Chang},
  {Chen}, {Ganguly}, {Jarvis}, {Kuo}, {Li}, {Liao}, {McLaughlin}, {Pen},
  {Peterson}, {Roman}, {Timbie}, {Voytek}, \& {Yadav}}]{mls+15}
{Masui}, K., {Lin}, H.-H., {Sievers}, J., {et~al.} 2015, Nature, 528, 523

\bibitem[{{Mazets} {et~al.}(2005){Mazets}, {Cline}, {Aptekar}, {Frederiks},
  {Golenetskii}, {Il'inskii}, \& {Pal'shin}}]{mca+05}
{Mazets}, E.~P., {Cline}, T.~L., {Aptekar}, R.~L., {et~al.} 2005, ArXiv
  Astrophysics e-prints, astro-ph/0502541

\bibitem[{{Meegan} {et~al.}(2009){Meegan}, {Lichti}, {Bhat}, {Bissaldi},
  {Briggs}, {Connaughton}, {Diehl}, {Fishman}, {Greiner}, {Hoover}, {van der
  Horst}, {von Kienlin}, {Kippen}, {Kouveliotou}, {McBreen}, {Paciesas},
  {Preece}, {Steinle}, {Wallace}, {Wilson}, \& {Wilson-Hodge}}]{mlb+09}
{Meegan}, C., {Lichti}, G., {Bhat}, P.~N., {et~al.} 2009, ApJ, 702, 791

\bibitem[{{Metzger} {et~al.}(2017){Metzger}, {Berger}, \& {Margalit}}]{mbm17}
{Metzger}, B.~D., {Berger}, E., \& {Margalit}, B. 2017, ApJ, 841, 14

\bibitem[{{Palmer} {et~al.}(2005){Palmer}, {Barthelmy}, {Gehrels}, {Kippen},
  {Cayton}, {Kouveliotou}, {Eichler}, {Wijers}, {Woods}, {Granot}, {Lyubarsky},
  {Ramirez-Ruiz}, {Barbier}, {Chester}, {Cummings}, {Fenimore}, {Finger},
  {Gaensler}, {Hullinger}, {Krimm}, {Markwardt}, {Nousek}, {Parsons}, {Patel},
  {Sakamoto}, {Sato}, {Suzuki}, \& {Tueller}}]{pbg+05}
{Palmer}, D.~M., {Barthelmy}, S., {Gehrels}, N., {et~al.} 2005, Nature, 434,
  1107

\bibitem[{{Paragi} {et~al.}(2012){Paragi}, {Shen}, {de Gasperin}, {Yang},
  {Merloni}, \& {Li}}]{psd+12}
{Paragi}, Z., {Shen}, Z.~Q., {de Gasperin}, F., {et~al.} 2012, in Proceedings
  of the 11th European VLBI Network Symposium and Users Meeting, ed.
  P.~{Charlot}, G.~{Bourda}, \& A.~{Collioud} (Trieste: SISSA), 8

\bibitem[{{Pen} \& {Connor}(2015)}]{pc15}
{Pen}, U.-L., \& {Connor}, L. 2015, ApJ, 807, 179

\bibitem[{{Petroff} {et~al.}(2016){Petroff}, {Barr}, {Jameson}, {Keane},
  {Bailes}, {Kramer}, {Morello}, {Tabbara}, \& {van Straten}}]{pbj+16}
{Petroff}, E., {Barr}, E.~D., {Jameson}, A., {et~al.} 2016, Proc. Astr. Soc.
  Aust., 33, e045

\bibitem[{{Popov} \& {Postnov}(2013)}]{pp13}
{Popov}, S.~B., \& {Postnov}, K.~A. 2013, ArXiv e-prints, arXiv:1307.4924

\bibitem[{Ransom(2001)}]{ran01}
Ransom, S.~M. 2001, PhD thesis, Harvard University

\bibitem[{{Scholz} {et~al.}(2016){Scholz}, {Spitler}, {Hessels}, {Chatterjee},
  {Cordes}, {Kaspi}, {Wharton}, {Bassa}, {Bogdanov}, {Camilo}, {Crawford},
  {Deneva}, {van Leeuwen}, {Lynch}, {Madsen}, {McLaughlin}, {Mickaliger},
  {Parent}, {Patel}, {Ransom}, {Seymour}, {Stairs}, {Stappers}, \&
  {Tendulkar}}]{ssh+16a}
{Scholz}, P., {Spitler}, L.~G., {Hessels}, J.~W.~T., {et~al.} 2016, ApJ, 833,
  177

\bibitem[{{Shannon} \& {Ravi}(2017)}]{sr17}
{Shannon}, R.~M., \& {Ravi}, V. 2017, ApJ, 837, L22

\bibitem[{{Spitler} {et~al.}(2014){Spitler}, {Cordes}, {Hessels}, {Lorimer},
  {McLaughlin}, {Chatterjee}, {Crawford}, {Deneva}, {Kaspi}, {Wharton},
  {Allen}, {Bogdanov}, {Brazier}, {Camilo}, {Freire}, {Jenet},
  {Karako-Argaman}, {Knispel}, {Lazarus}, {Lee}, {van Leeuwen}, {Lynch},
  {Ransom}, {Scholz}, {Siemens}, {Stairs}, {Stovall}, {Swiggum},
  {Venkataraman}, {Zhu}, {Aulbert}, \& {Fehrmann}}]{sch+14}
{Spitler}, L.~G., {Cordes}, J.~M., {Hessels}, J.~W.~T., {et~al.} 2014, ApJ,
  790, 101

\bibitem[{{Spitler} {et~al.}(2016){Spitler}, {Scholz}, {Hessels}, {Bogdanov},
  {Brazier}, {Camilo}, {Chatterjee}, {Cordes}, {Crawford}, {Deneva}, {Ferdman},
  {Freire}, {Kaspi}, {Lazarus}, {Lynch}, {Madsen}, {McLaughlin}, {Patel},
  {Ransom}, {Seymour}, {Stairs}, {Stappers}, {van Leeuwen}, \& {Zhu}}]{ssh+16}
{Spitler}, L.~G., {Scholz}, P., {Hessels}, J.~W.~T., {et~al.} 2016, Nature,
  531, 202

\bibitem[{{Starling} {et~al.}(2013){Starling}, {Willingale}, {Tanvir}, {Scott},
  {Wiersema}, {O'Brien}, {Levan}, \& {Stewart}}]{swt+13}
{Starling}, R.~L.~C., {Willingale}, R., {Tanvir}, N.~R., {et~al.} 2013, MNRAS,
  431, 3159

\bibitem[{{Tendulkar} {et~al.}(2016){Tendulkar}, {Kaspi}, \& {Patel}}]{tkp16}
{Tendulkar}, S.~P., {Kaspi}, V.~M., \& {Patel}, C. 2016, ApJ, 827, 59

\bibitem[{{Tendulkar} {et~al.}(2017){Tendulkar}, {Bassa}, {Cordes}, {Bower},
  {Law}, {Chatterjee}, {Adams}, {Bogdanov}, {Burke-Spolaor}, {Butler},
  {Demorest}, {Hessels}, {Kaspi}, {Lazio}, {Maddox}, {Marcote}, {McLaughlin},
  {Paragi}, {Ransom}, {Scholz}, {Seymour}, {Spitler}, {van Langevelde}, \&
  {Wharton}}]{tbc+17}
{Tendulkar}, S.~P., {Bassa}, C.~G., {Cordes}, J.~M., {et~al.} 2017, ApJ, 834,
  L7

\bibitem[{{Terashima} \& {Wilson}(2003)}]{tw03}
{Terashima}, Y., \& {Wilson}, A.~S. 2003, ApJ, 583, 145

\bibitem[{{Thornton} {et~al.}(2013){Thornton}, {Stappers}, {Bailes},
  {Barsdell}, {Bates}, {Bhat}, {Burgay}, {Burke-Spolaor}, {Champion}, {Coster},
  {D'Amico}, {Jameson}, {Johnston}, {Keith}, {Kramer}, {Levin}, {Milia}, {Ng},
  {Possenti}, \& {van Straten}}]{tsb+13}
{Thornton}, D., {Stappers}, B., {Bailes}, M., {et~al.} 2013, Science, 341, 53

\bibitem[{{Verner} {et~al.}(1996){Verner}, {Ferland}, {Korista}, \&
  {Yakovlev}}]{vfky96}
{Verner}, D.~A., {Ferland}, G.~J., {Korista}, K.~T., \& {Yakovlev}, D.~G. 1996,
  ApJ, 465, 487

\bibitem[{{Wilms} {et~al.}(2000){Wilms}, {Allen}, \& {McCray}}]{wam00}
{Wilms}, J., {Allen}, A., \& {McCray}, R. 2000, ApJ, 542, 914

\bibitem[{{Yao} {et~al.}(2017){Yao}, {Manchester}, \& {Wang}}]{ymw17}
{Yao}, J.~M., {Manchester}, R.~N., \& {Wang}, N. 2017, ApJ, 835, 29

\bibitem[{{Younes} {et~al.}(2016){Younes}, {Kouveliotou}, {Huppenkothen},
  {Gogus}, {Kaneko}, \& {van der Horst}}]{ykh+16}
{Younes}, G., {Kouveliotou}, C., {Huppenkothen}, D., {et~al.} 2016, The
  Astronomer's Telegram, 8781

\end{thebibliography}

\end{document}